\documentclass{article}
\usepackage{spconf,amsmath,graphicx}
\usepackage{lipsum, soul,xcolor} 
\usepackage{hyperref}
\usepackage{amsfonts}

\let\OLDthebibliography\thebibliography
\renewcommand\thebibliography[1]{
	\OLDthebibliography{#1}
	\setlength{\parskip}{0pt}
	\setlength{\itemsep}{0pt plus 0.3ex}
}

\title{ADL-MVDR: All deep learning MVDR beamformer for target speech separation}
\name{Zhuohuang Zhang$^{1,2\star}$, Yong Xu$^{2}$, Meng Yu$^{2}$, Shi-Xiong Zhang$^{2}$, Lianwu Chen$^{2}$, Dong Yu$^{2}$
\thanks{This work was done while Z. Zhang was a research intern at Tencent AI Lab, Bellevue, USA. $\star$ \href{mailto: zhuozhan@iu.edu}{zhuozhan@iu.edu}}
}
\address{
$^{1}$ Indiana University, Bloomington, USA \quad $^{2}$ Tencent AI Lab\\
}

\begin{document}
%
\maketitle

\begin{abstract}
Speech separation algorithms are often used to separate the target speech from other interfering sources. However, purely neural network based speech separation systems often cause nonlinear distortion that is harmful for automatic speech recognition (ASR) systems. The conventional mask-based minimum variance distortionless response (MVDR) beamformer can be used to minimize the distortion, but comes with high level of residual noise. Furthermore, the matrix operations (e.g., matrix inversion) involved in the conventional MVDR solution are sometimes numerically unstable when jointly trained with neural networks. In this paper, we propose a novel all deep learning MVDR framework, where the matrix inversion and eigenvalue decomposition are replaced by two recurrent neural networks (RNNs), to resolve both issues at the same time. The proposed method can greatly reduce the residual noise while keeping the target speech undistorted by leveraging on the RNN-predicted frame-wise beamforming weights. The system is evaluated on a Mandarin audio-visual corpus and compared against several state-of-the-art (SOTA) speech separation systems. Experimental results demonstrate the superiority of the proposed method across several objective metrics and ASR accuracy. 
\end{abstract}
\begin{keywords}
Speech separation, speech enhancement, MVDR, ADL-MVDR, deep learning
\end{keywords}
%

\section{Introduction}
\label{sec:intro}

Environmental noises and adverse room acoustics can greatly affect the quality of the speech signal and therefore degrade the effectiveness of many speech communication systems (e.g., digital hearing-aid devices \cite{van2009speech}, and automatic speech recognition (ASR) systems \cite{du2014robust,weninger2015speech,AVJW2020}). Speech enhancement and speech separation algorithms are thus proposed to alleviate this problem. With the renaissance of neural networks, better objective performance can be achieved using deep learning methods \cite{wang2014training,luo2018tasnet,zhang2020loss}. However, it often results in greater amount of nonlinear distortion on the separated target speech \cite{xu2020neural,luo2019conv, tan2020audio}, which harms the performance of ASR systems. 

The minimum variance distortionless response (MVDR) filters \cite{van1988beamforming} aim to reduce the noise while keeping the target speech undistorted. More recently, MVDR systems with neural network (NN) based time-frequency (T-F) mask estimator can help greatly reduce the word error rate (WER) of ASR systems with less amount of distortion \cite{heymann2016neural, erdogan2016improved, xu2019joint}, yet they still suffer from residual noise problems since chunk- or utterance-level beamforming weights \cite{xiao2017time,boeddeker2018exploring,xu2020neural} are not optimal for noise reduction. Some frame-level MVDR weights estimation methods have been proposed, in \cite{souden2011integrated}, the authors estimate the covariance matrix in a recursive way. Nevertheless, the matrix inversion involved in the MVDR solution is sometimes numerically unstable~\cite{zhao2012fast} when jointly trained with NNs, where techniques such as diagonal loading is often used to alleviate this issue~\cite{mestre2003diagonal}. Prior studies have indicated that it is feasible for a recurrent neural network (RNN) to learn the matrix inversion efficiently \cite{wang1993recurrent, zhang2005design} and that RNNs can better stabilize the process of matrix inversion and principal component analysis (PCA) when jointly trained with NNs. 

There are three main contributions in this work, firstly, we propose a novel all deep learning MVDR framework (denoted as ADL-MVDR) where the ADL-MVDR can be jointly trained stably with the front-end filter estimator for predicting frame-level beamforming weights. Secondly, we propose to use two RNNs to replace the matrix inversion and PCA involved in the MVDR solution, instead of utilizing the traditional mathematical approach. Thirdly, instead of using the classical per T-F bin mask, we adopt a complex ratio filtering method \cite{mack2019deep} (denoted as cRF) to further stabilize joint training process and estimate the covariance matrices of target speech and noise more accurately. The RNN components of ADL-MVDR system help to recursively estimate the statistical variables (i.e., inverse of the noise covariance matrix and PCA of the steering vector) in an adaptive way. 
Meanwhile, a Conv-TasNet variant \cite{luo2019conv, tan2020audio} is adopted as the front-end filter estimator to calculate the frame-level covariance matrices. 

The proposed cRF based ADL-MVDR system achieves the best performance in many objective metrics as well as the ASR accuracy. To the best of our knowledge, this is the first pioneering study that applies RNNs to derive the MVDR solution by replacing the matrix inversion and PCA. Note that Xiao et al. \cite{xiao2016study} once proposed a directly NN-learned beamforming weights method which was not successful due to lack of using noise information, whereas our approach still follows the mask-based MVDR framework and explicitly utilizes the noise and speech covariance matrices with RNNs.

The rest of the paper is organized as follows: Section \ref{sec:maskMVDR} introduces the conventional mask-based MVDR beamformer and Section \ref{sec:neuralMVDR} describes the proposed ADL-MVDR beamformer. We present the dataset and experimental setup in Section \ref{sec:expsetup}. Results are reported in Section \ref{sec:results}. Finally, we draw conclusions in Section \ref{sec:conclusions}.




\vspace{-2mm}
\section{Signal model for MVDR Beamformer}
\label{sec:maskMVDR}
\vspace{-1mm}

This section describes the conventional mask-based MVDR beamformer, the proposed ADL-MVDR beamformer will be introduced in the next section. Consider a noisy speech mixture $\mathbf{y = [y_1, y_2, ..., y_M]^T}$ recorded with an $M$-size microphone array. Let $\mathbf{s}$ represent the clean speech and let $\mathbf{n}$ denote the interfering noise with $M$ channels, then we have 
\begin{equation}
\label{noisy mixture}
\mathbf{Y}(t,f) = \mathbf{S}(t,f) + \mathbf{N}(t,f), 
\end{equation}
where $(t,f)$ indicates the time and frequency indices of the acoustic signals in the T-F domain, and $\mathbf{Y}, \mathbf{S}, \mathbf{N}$ denote the corresponding variables in T-F domain. The separated speech $\mathbf{\hat{s}_{MVDR}}(t,f)$ can be obtained as
\begin{equation}
\label{beamforming}
\mathbf{\hat{s}_{MVDR}}(t,f) = \mathbf{h}^{\mathrm{H}}(f)\mathbf{Y}(t,f),
\end{equation}
where $\mathbf{h}(f)\in \mathbb{C}^{M}$ represents the MVDR weights at frequency index $f$ and $^{\mathrm{H}}$ stands for the Hermitian operator. The goal of the MVDR beamformer is to minimize the power of the noise while keeping the target speech undistorted, which can be formulated as
\begin{equation}
\mathbf{h_{MVDR}}=\underset{\mathbf{h}}{\arg \min \mathbf{h}}^{\mathrm{H}} \mathbf{\Phi}_{\text{NN}} \mathbf{h} \quad \bf{\text {s.t.}} \quad \mathbf{h}^{\mathrm{H}} \mathbf{\boldsymbol{v}}=\mathbf{1},
\end{equation}
here $\mathbf{\Phi}_{\text{NN}}$ stands for the covariance matrix of the noise power spectral density (PSD) and $\boldsymbol{v}(f) \in \mathbb{C}^{M}$ denotes the steering vector of the target speech. Different solutions can be used to derive the MVDR beamforming weights. In our study, we mainly focus on the MVDR solution that is based on the steering vector \cite{higuchi2018frame,shimada2018unsupervised}, which can be derived by applying PCA on the speech covariance matrix.
\begin{equation}
\label{MVDR_solutions}
\mathbf{h}(f) =\frac{\mathbf{\Phi}_{\text{NN}}^{-1}(f) \boldsymbol{v}(f)}{\mathbf{\boldsymbol{v}}^{\mathrm{H}}(f) \mathbf{\Phi}_{\text{NN}}^{-1}(f) \mathbf{\boldsymbol{v}}(f)}, \quad \mathbf{h}(f) \in \mathbb{C}^{M}, \\
\end{equation}
note that we use diagonal loading~\cite{mestre2003diagonal} to alleviate the numerical instability issue in the matrix inversion~\cite{zhao2012fast}.


A complex ratio mask \cite{williamson2015complex} (denoted as cRM) can be used to estimate the target speech accurately with less amount of phase distortion, which benefits human listeners \cite{williamson2015complex,zhang2020investigation}. In this case, the estimated speech $\mathbf{\hat{S}}_\text{cRM}$ and covariance matrix of the speech PSD $\mathbf{\Phi}_{\text{SS}}$ can be computed as
\vspace{-2mm}
\begin{equation}
\begin{aligned}
\mathbf{\hat{S}}_\mathrm{\text{cRM}}(t,f) &= \mathrm{\mathbf{M}}_{\mathrm{S}}(t,f) * \mathbf{Y}(t,f), \\
\mathbf{\Phi}_{\text{SS}}(f) &= \frac{\sum_{t=1}^{T} \mathbf{\hat{S}}_\mathrm{\text{cRM}}(t,f) \mathbf{\hat{S}}_\mathrm{\text{cRM}}^{\mathrm{H}}(t,f)}{\sum_{t=1}^{T} \mathrm{\mathbf{M}}_{\text{S}}^{\mathrm{H}}(t,f) \mathrm{\mathbf{M}}_{\text{S}}(t,f)}, 
\end{aligned}
\end{equation}
where $*$ denotes the complex multiplier and $\mathrm{\mathbf{M}}_{\text{S}}$ represents the estimated cRM for speech target. The noise covariance matrix $\mathbf{\Phi}_{\text{NN}}$ can be obtained in a similar way. Note that the covariance matrix $\mathbf{\Phi}$ derived here is on the utterance-level, leading to utterance-level beamforming weights which are not optimal for noise reduction on each frame.

\vspace{-3mm}
\section{Proposed ADL-MVDR Beamformer}
\label{sec:neuralMVDR}
\vspace{-1mm}

\begin{figure*}[htb!]
  \centering
  \includegraphics[scale = 0.35]{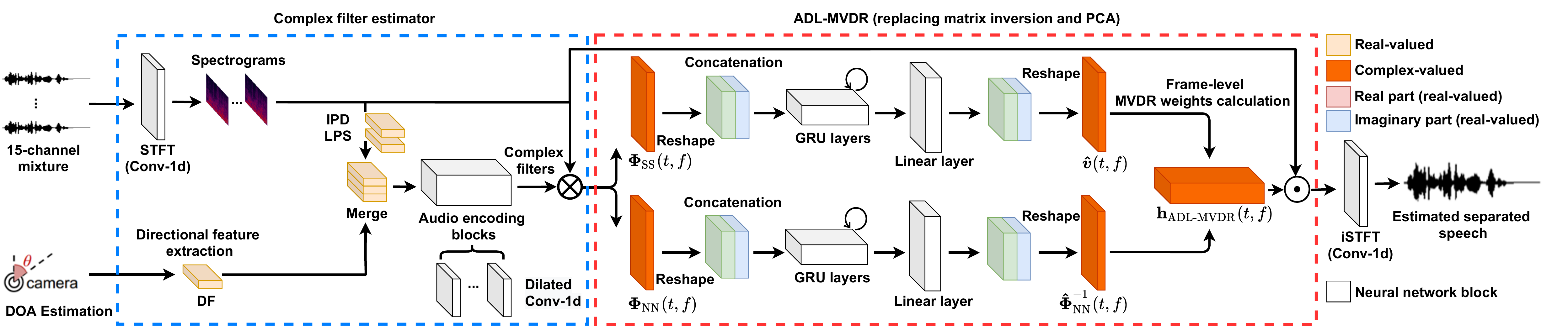}
  \vspace{-6.5mm}
  \caption{Network structure of proposed ADL-MVDR beamformer. $\otimes$ and $\odot$ indicate the operations expressed in Eq. (\ref{eq:cRF_filtering}) and (\ref{tf-beamforming}), respectively. The complex filter estimation (i.e., cRF) and ADL-MVDR (i.e., estimations of $\mathbf{\boldsymbol{v}}(t,f)$ and $\mathbf{\Phi}_{\text{NN}}^{-1}(t,f)$) blocks are highlighted in the blue and red dashed boxes, respectively. The real and imaginary parts are reshaped and concatenated before fed into the GRU networks, and then reshaped again as inputs for calculating MVDR weights. The estimated frame-level MVDR weights are then applied to the multi-channel speech.}
  \label{fig:GRU_MVDR}
  \vspace{-5mm}
\end{figure*}

In this work, we implement two gated recurrent unit (GRU) \cite{chung2014empirical} based networks (denoted as GRU-Nets) to replace the matrix inversion and PCA in Eq. (\ref{MVDR_solutions}) for estimating frame-level beamforming weights. One advantage of using RNNs is that it utilizes the weighted information from all previous frames and does not need any heuristic updating factors between consecutive frames as needed in recursive approaches \cite{souden2011integrated,tammen2019dnn}.

\vspace{-3mm}
\subsection{cRF for covariance matrix estimation}
To better utilize the nearby T-F information and stabilize the estimated statistical variables (namely, $\mathbf{\Phi}_{\text{SS}}(t,f)$ and $\mathbf{\Phi}_{\text{NN}}(t,f)$), we adopt a cRF method \cite{mack2019deep} to estimate the speech and noise components. For each T-F bin, the cRF is applied to its $(2K+1) \times (2L+1)$ nearby T-F bins as
\vspace{-3.5mm}
\begin{equation}
\begin{aligned}
\label{eq:cRF_filtering}
\mathbf{\hat{S}}_\mathrm{\text{cRF}}(t,f) &= \sum_{\tau_{1}=-L}^{L}\sum_{\tau_{2}=-K}^{K}\mathrm{\mathbf{F}_{\text{S}}}(t+\tau_{1},f+\tau_{2})\\
&*\mathbf{Y}(t+\tau_{1},f+\tau_{2}), \\
\mathbf{\Phi}_{\text{SS}}(t,f) &=\frac{\mathbf{\hat{S}}_\text{cRF}(t,f) \mathbf{\hat{S}}_\text{cRF}^{\mathrm{H}}(t,f)}{\sum_{t=1}^{T}\mathbf{M}_{\mathrm{S}}^{\mathrm{H}}(t,f) \mathbf{M}_{\mathrm{S}}(t,f)}, 
\end{aligned}
\vspace{-3mm}
\end{equation}
here $\mathbf{\hat{S}}_\mathrm{\text{cRF}}$ indicates the estimated speech using the speech complex ratio filter, i.e., $\mathrm{\mathbf{F}_{\text{S}}}$. The cRF is equivalent to $(2K+1) \times (2L+1)$ number of cRMs that each applies to the corresponding shifted version (i.e., along time and frequency axes) of the noisy spectrogram. The frame-level speech covariance matrix is then computed where the center mask of the cRF (i.e., $\mathbf{M}_{\mathrm{S}}(t,f)$) is used for normalization. Note that we do not sum over the time dimension of $\mathbf{\Phi}_{\text{SS}}$ in order to preserve the frame-level temporal information. The frame-level noise covariance matrix $\mathbf{\Phi}_{\text{NN}}(t,f)$ can be obtained in a similar way.

\vspace{-2mm}
\subsection{RNNs for replacing matrix inversion and PCA in MVDR}
Here we propose to replace the mathematical derivation of the steering vector and the inverse of noise covariance matrix with two GRU-Nets. The GRU-Nets can better utilize temporal information from previous frames for estimating statistical terms than conventional frame-wise approaches that are based on heuristic updating factors \cite{souden2011integrated,tammen2019dnn}. Additionally, replacing the matrix inversion with a GRU-Net resolves the instability issue during joint training with NNs. We conjecture that these GRU-Nets will learn the steering vector and the matrix inversion through back propagation, i.e.,
\begin{equation}
\begin{aligned}
\mathbf{\hat{\boldsymbol{v}}}(t,f) & = \mathbf{GRU{\text -}Net}_{\boldsymbol{v}}(\mathbf{\Phi}_{\text{SS}}(t,f)), \\
\mathbf{\hat{\Phi}}_{\text{NN}}^{-1}(t,f) & = \mathbf{GRU{\text -}Net}_{\text{NN}}(\mathbf{\Phi}_{\text{NN}}(t,f)), \\
\end{aligned}
\vspace{-2mm}
\end{equation}
where the real and imaginary parts of the complex-valued covariance matrix $\mathbf{\Phi}$ are concatenated together as input to the GRU-Net. Here we assume that the explicitly calculated speech and noise covariance matrices are important for RNNs to learn the spatial filtering, which is different from the directly NN-learned beamforming weights in \cite{xiao2016study}. Leveraging on the temporal structure of RNNs, the model recursively accumulates and updates the covariance matrix for each frame. As shown in Fig. \ref{fig:GRU_MVDR}, the output of each GRU-Net is fed into a linear layer to obtain the final real and imaginary parts of the complex-valued covariance matrix or steering vector. Then, we compute the frame-level ADL-MVDR weights as
\begin{equation}
\begin{aligned}
\label{MVDR_tf_solutions}
\mathbf{h}(t,f) & = \frac{\mathbf{\hat{\Phi}}_{\text{NN}}^{-1}(t,f) \hat{\boldsymbol{v}}(t,f)}{\mathbf{\hat{\boldsymbol{v}}^{\mathrm{H}}}(t,f) \mathbf{\hat{\Phi}}_{\text{NN}}^{-1}(t,f) \mathbf{\hat{\boldsymbol{v}}}(t,f)}, \quad  \mathbf{h}(t,f) \in \mathbb{C}^{M}.
\end{aligned}
\end{equation}
Here $\mathbf{h}(t,f)$ is frame-wise and different from the utterance-level weights of conventional mask-based MVDR defined in Eq. (\ref{MVDR_solutions}). Finally, the ADL-MVDR enhanced speech is obtained
\vspace{-2mm}
\begin{equation}
\label{tf-beamforming}
\mathbf{\hat{S}_{ADL{\text -}MVDR}}(t,f) = \mathbf{h}^{\mathrm{H}}(t,f)\mathbf{Y}(t,f).
\end{equation}

\vspace{-4.5mm}
\section{Dataset and Experimental Setup}
\label{sec:expsetup}
\vspace{-2mm}


\begin{table*}[t!]
\centering
\caption{Experimental results for different speech separation systems across objective evaluation metrics.}
\scalebox{0.79}{
\begin{tabular}{l|cccc|ccc|c|c|c|c}
\hline
\multicolumn{1}{c|}{Systems/Metrics} & \multicolumn{8}{c|}{PESQ $\in [-0.5,4.5]$} & Si-SNR (dB) & \multicolumn{1}{l|}{SDR (dB)} & WER ($\%$) \\ \hline
 & 0-15$^{\circ}$ & 15-45$^{\circ}$ & 45-90$^{\circ}$ & 90-180$^{\circ}$ & 1spk & 2spk & 3spk & Avg. & Avg. & Avg. & Avg. \\ \hline
Reverberant clean (reference) & 4.50 & 4.50 & 4.50 & 4.50 & 4.50 & 4.50 & 4.50 & 4.50 & $\infty$ & $\infty$ & 8.26 \\
Noisy Mixture (interfering speech + noise) & 1.88 & 1.88 & 1.98 & 2.03 & 3.55 & 2.02 & 1.77 & 2.16 & 3.39 & 3.50 & 55.14 \\ \hline
NN with cRM & 2.72 & 2.92 & 3.09 & 3.07 & 3.96 & 3.02 & 2.74 & 3.07 & 12.23 & 12.73 & 22.49 \\
NN with cRF (3$\times$3) & 2.75 & 2.95 & 3.12 & 3.09 & 3.98 & 3.06 & 2.76 & 3.10 & 12.50 & 13.01 & 22.07 \\ \hline
MVDR with cRM \cite{xu2020neural} & 2.55 & 2.76 & 2.96 & 2.84 & 3.73 & 2.88 & 2.56 & 2.90 & 10.62 & 12.04 & 16.85\\ 
MVDR with cRF (3$\times$3) & 2.55 & 2.77 & 2.96 & 2.89 & 3.82 & 2.90 & 2.55 & 2.92 & 11.31 & 12.58 & 15.91\\ 
Multi-tap MVDR with cRM (2-tap) \cite{xu2020neural} & 2.70 & 2.96 & 3.18 & 3.09 & 3.80 & 3.07 & 2.74 & 3.08 & 12.56 & 14.11 & 13.67 \\ 
Multi-tap MVDR with cRF (2-tap, 3$\times$3) & 2.67 & 2.95 & 3.15 & 3.10 & 3.92 & 3.06 & 2.72 & 3.08 & 12.66 & 14.04 & 13.52 \\ \hline
\textbf{Proposed ADL-MVDR with cRF (3$\times$3)} & $\textbf{3.04}$ & $\textbf{3.30}$ & $\textbf{3.48}$ & $\textbf{3.48}$ & $\textbf{4.17}$ & $\textbf{3.41}$ & $\textbf{3.07}$ & $\textbf{3.42}$ & $\textbf{14.80}$ & $\textbf{15.45}$ & $\textbf{12.73}$ \\ \hline
\end{tabular}
}
\vspace{-5mm}
\end{table*}

\subsection{System overview and dataset}
\vspace{-1mm}
The proposed system is evaluated based on our previously reported multi-modal multi-channel target speech separation platform \cite{tan2020audio,gu2020multi}. As shown in Fig. \ref{fig:GRU_MVDR}, we extract the log-power interaural phase difference (IPD) and spectra (LPS) features from the 15-channel microphone recorded mixture that is synchronized with the $180^{\circ}$ camera \cite{tan2020audio}. The direction of arrival (DOA) is roughly estimated using the location of the target speaker's face in the whole camera view \cite{tan2020audio}, then the location guided directional feature (DF) \cite{chen2018multi} is extracted. The DF is then merged with the IPD and LPS features before fed into the audio encoding blocks \cite{tan2020audio,gu2020multi}. We use our previously reported Mandarin audio-visual dataset \cite{xu2020neural,tan2020audio,gu2020multi} collected from Youtube as the speech corpus (will be released \cite{zhang2020multi} soon). Different from our previous works \cite{tan2020audio,gu2020multi}, the lip movement feature is not fed into the model in this study as we focus on the beamforming. The corpus contains 205500 audio clips (roughly 200 hours) with sampling rate set to 16 kHz. The simulated multi-channel audio data contains sources from different speakers (either target or interfering sources). The audios are further mixed with random cuts of noises recorded indoors and different reverberation conditions (T60s from 0.05 s to 0.7 s) are applied \cite{tan2020audio}.


\vspace{-4mm}
\subsection{Experimental setup}
\vspace{-1mm}


A 512-point FFT is applied along with 32 ms Hann window and 16 ms step size to extract the audio features. The size of the cRF is empirically set to 3$\times$3 (i.e., K and L in Eq. (\ref{eq:cRF_filtering}) set to 1). In the training stage, the audio chunk and batch size are set to 4 s and 12, respectively. Adam optimizer \cite{kingma2014adam} is adopted with the initial learning rate set to 1e$^{-3}$. The objective is to maximize the time-domain scale-invariant source-to-noise ratio (Si-SNR) \cite{luo2019conv}. All models are trained for 60 epochs.

In terms of the GRU-Nets, the $\mathbf{\hat{\boldsymbol{v}}}(t,f)$ estimation network consists of two layers of GRU followed by another layer of fully connected (FC) neurons. The hidden size is set to 500 and 250 for the 2-layer GRU with tanh activation function, linear activation function is used for the FC layer with a hidden size of 30. As for $\mathbf{\hat{\Phi}}_{\text{NN}}^{-1}(t,f)$ estimation, the corresponding GRU-Net features a similar structure, where each GRU layer contains 500 units with a 450-size FC layer. 

Four baseline systems are considered, including a purely NN-based (i.e., a Conv-TasNet variant \cite{luo2019conv}) cRM system \cite{xu2020neural}, a purely NN-based cRF system, a conventional cRM-based MVDR system \cite{xu2020neural} and another cRF-based MVDR system (denoted as NN with cRM, NN with cRF, MVDR with cRM, and MVDR with cRF, respectively). We further include two multi-tap (i.e., $[t, t-1]$) MVDR systems that are proposed in our previous work \cite{xu2020neural}, trained with cRM and cRF (denoted as Multi-tap MVDR with cRM/cRF). 

\vspace{-3.5mm}
\section{Results and discussions}
\label{sec:results}
\vspace{-2mm}

The systems' performance \footnote{Demos (including real-world recording evaluations) are available at \href{https://zzhang68.github.io/adlmvdr/}{https://zzhang68.github.io/adlmvdr/}} is evaluated by several objective metrics, including PESQ~\cite{rix2001perceptual}, Si-SNR \cite{luo2019conv}, and signal-to-distortion ratio (SDR)~\cite{vincent2006performance}. A Tencent commercial mandarin speech recognition API is used for measuring the WER in this study. The PESQ scores are further presented in specific conditions (i.e., the angle between the target speaker and the closest interfering source, and the number of speakers). Average scores for other metrics are presented. Note that the systems are only trained on speech separation and denoising, without dereverberation.

\begin{figure}[t!]
  \centering
  \includegraphics[scale = 0.35]{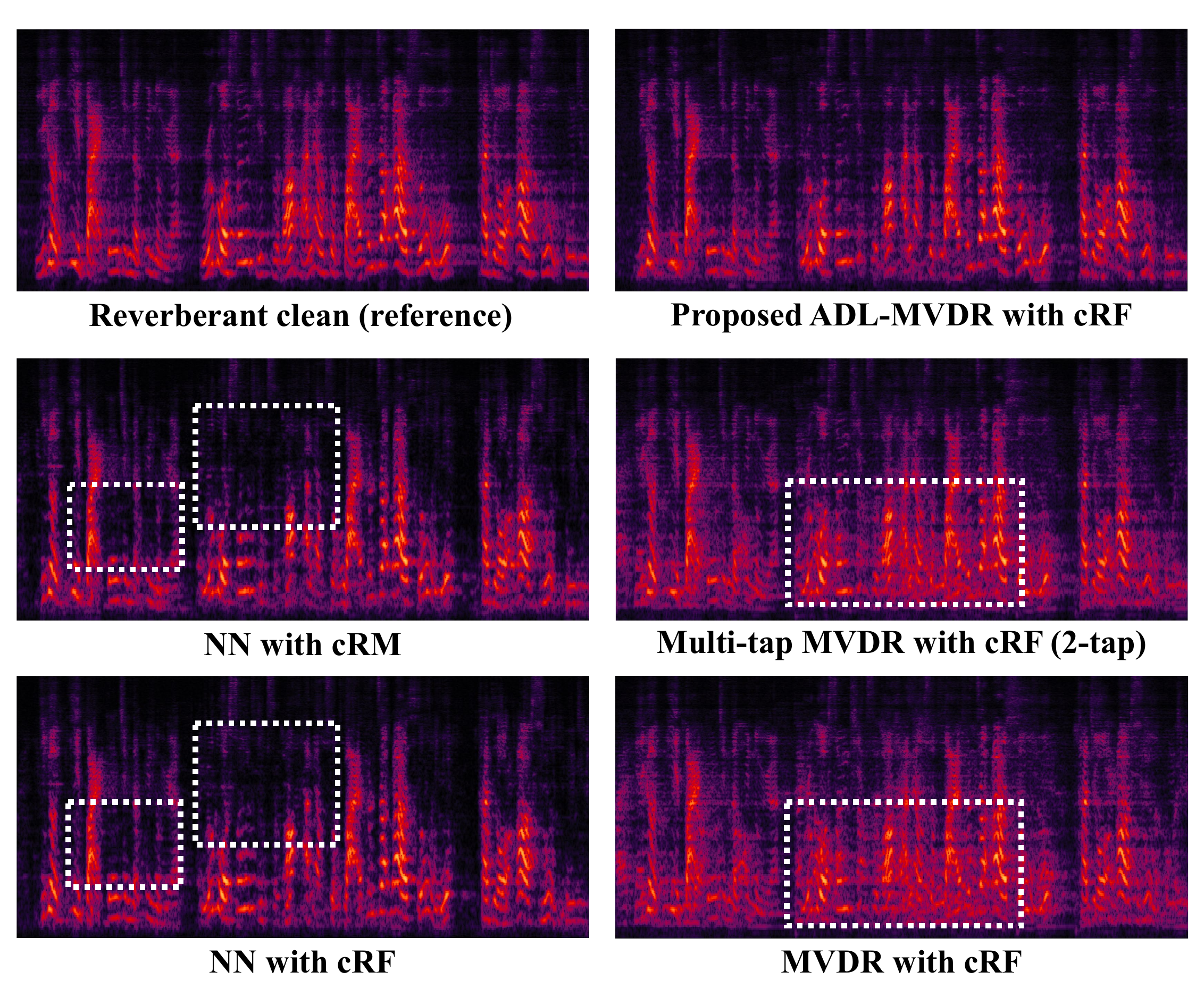}
  \vspace{-4mm}
  \caption{Sample spectrograms of some evaluated systems}
  \label{fig:sample_spectrogram}
  \vspace{-4mm}
\end{figure}

$\textbf{ADL-MVDR vs. NN}$: the proposed ADL-MVDR system achieves significantly better results across all metrics and ASR accuracy than purely NN-based systems. Our proposed ADL-MVDR system achieves around 42\% improvement on WER (i.e., 12.73\% vs. 22.07\%) when compared to NN with cRF. Significant improvements aross objective metrics are also observed (i.e., PESQ: 3.42 vs. 3.10, Si-SNR: 14.80 dB vs. 12.50 dB, SDR: 15.45 dB vs. 13.01 dB). For purely NN-based systems, although they perform reasonably well across objective metrics, they perform poorly in ASR system due to large amount of distortion (also highlighted in Fig. \ref{fig:sample_spectrogram}). 

$\textbf{ADL-MVDR vs. MVDR}$: our proposed ADL-MVDR system achieves about 17\% PESQ improvement over the baseline MVDR system with cRF (i.e., 3.42 vs. 2.92). In terms of ASR accuracy, the proposed ADL-MVDR system outperforms MVDR with cRF by a large margin (i.e., 12.73\% vs. 15.91\%). Considering that the commercial ASR system is already robust to some mild noise, the differences on WER become smaller for multi-tap MVDR systems, yet large gaps can be found in all other metrics (e.g., 0.34 absolute improvement on average PESQ). Although conventional MVDR systems can alleviate the distortion issue, there still remains a lot of residual noise. This can be observed in the objective scores and is also highlighted in Fig. \ref{fig:sample_spectrogram}. Again, our proposed ADL-MVDR system resolves this issue (i.e., Si-SNR: 14.80 dB vs. 12.66 dB and SDR: 15.45 dB vs. 14.04 dB) while also keeping the target speech undistorted. Specifically, under extreme conditions where interfering sources are very close to each other (e.g., angles between 0-15$^{\circ}$), our proposed ADL-MVDR system improves the speech quality by nearly 62\% (i.e., PESQ: 3.04 vs. 1.88). The experimental results presented here verify our claims that the proposed ADL-MVDR system not only ensures the distortionless of the target speech (i.e., lowest WER) but also eliminates the residual noise (i.e., highest scores across all objective metrics).

$\textbf{cRM vs. cRF}$: the NN with cRF achieves better performance in all metrics (e.g., Si-SNR: 12.50 dB vs. 12.23 dB) and ASR accuracy (i.e., 22.07\% vs. 22.49\%) than NN with cRM. Slight improvements can be found on conventional MVDR systems due to utterance-level weights. The cRF is more important for ADL-MVDR system since ADL-MVDR is recursively getting frame-level weights from the estimated covariance matrices. It indicates that the benefits of introducing T-F filtering include further reducing the residual noise while not distorting the speech. 



\vspace{-3mm}
\section{Conclusions and future work}
\label{sec:conclusions}
\vspace{-2mm}

In this paper, we proposed a novel all deep learning MVDR method to recursively learn the spatio-temporal filtering for multi-channel target speech separation. The proposed system outperforms prior arts across several objective metrics and ASR accuracy. The future of our proposed ADL-MVDR framework is promising and it could be generalized to many speech separation systems. We will further verify this idea on single-channel speech separation and dereverberation tasks.

\bibliographystyle{IEEEbib}
\bibliography{refs}

\end{document}